**Abstract:** Radiology has been essential to accurately diagnosing diseases and assessing responses to treatment. The challenge however lies in the shortage of radiologists globally. As a response to this, a number of Artificial Intelligence solutions are being developed. The challenge Artificial Intelligence radiological solutions however face is the lack of a benchmarking and evaluation standard, and the difficulties of collecting diverse data to truly assess the ability of such systems to generalise and properly handle edge cases. We are proposing a radiograph-agnostic platform and framework that would allow any Artificial Intelligence radiological solution to be assessed on its ability to generalise across diverse geographical location, gender and age groups.

**Overview**

An estimated 3.6 billion diagnostic medical examinations, such as X-rays, are performed worldwide every year. Advances in radiology technology has improved illness and injury diagnosis and treatments. These radiological procedures include X-Rays, Mammograms, Ultrasound, PET (positron emission tomography) scans, MRI (magnetic resonance imaging) scans and CT (computed tomography) scans. They are used mainly in dealing with a broad range of non-communicable or chronic diseases. These are primarily cardiovascular diseases, cancer, chronic respiratory diseases and diabetes. Radiology has helped in the rapid non-invasive screening of conditions such as breast cancer, which reduces the mortality rate, especially with early detection. 33 million screening mammography exams are performed each year in the United States alone. Research led by led by Elizabeth Kagan Arleo, MD, of Weill Cornell Medicine found that recommendation of annual screening starting at age 40 would result in a nearly 40 percent reduction in deaths due to breast cancer (Arleo et al, 2017). Simple radiological procedures like ultrasound can reduce the need for surgical interventions. And though clinical judgement may be sufficient, radiological procedures are necessary in confirming and properly evaluating the causes of many conditions and responses to treatments.

*Challenges Facing Radiology*

Though radiology is very important, there's a shortage of radiologists globally, especially in developing countries. Liberia, for example, only has about 2 radiologists (RAD-AID, 2017), whilst Ghana has 34 radiologists and Kenya has 200 radiologists (UCSF, 2015). And in the UK, only one-in-five trusts and health boards has sufficient number of interventional radiologists to run a safe 24/7 service to perform urgent procedures (Clinical Radiology UK Workforce Census Report, 2018) whilst their workload of reading and interpreting medical images has increased by 30% between 2012 and 2017. There's a need for scalable and accurate automated radiological systems. Deep Learning, especially Convolutional Neural Networks, is gaining wide attention for its ability to accurately analyse medical images, with the potential to help solve the shortage of radiologists.

*Artificial Intelligence In Radiology*

The re-emergence of Artificial Intelligence (A.I) and Deep Learning, due to growth in computing power and data, has led to advancements in Deep Convolutional Neural Networks, which has allowed for breakthrough research and applications in Radiology. Artificial Intelligence and Deep Learning holds a lot of potential in Radiology. Artificial Intelligence can provide support to radiologists and alleviate radiologist fatigue. It can help in flagging patients who require urgent care to radiologists and physicians. Deep Learning could also help increase interrater reliability among radiologists throughout their years in clinical practice. A recent study found that the Fleiss' kappa measure of interrater reliability for detecting anterior cruciate ligament tear, meniscal tear, and abnormality were higher with model assistance than without it (Bien et al., 2018). Deep Learning has achieved performances comparable to humans and sometimes better. A recent study analysed 14 research works done using Deep Learning to detect diseases via medical images, they found that on average, Deep Learning systems correctly detected a disease state 87% of the time – compared with 86% for healthcare professionals – and correctly gave the all-clear 93% of the time, compared with 91% for human experts (Liu et al., 2019). Deep Learning has performed as well as radiologists and sometimes better at detecting abnormalities like pneumonia, fibrosis, hernia, edema and pneumothorax in chest x-rays (Rajpurkar et. al, 2017). It has also been used to detect knee abnormalities via magnetic resonance (MR) imaging at near-human-level performance (Bien et. al, 2018). Researchers have also trained Deep Learning models that outperformed dermatologists at detecting skin cancer (Esteva et. al, 2017, Haenssle et. al, 2018).

*Research Data*

One key focus of deep learning radiological applications is breast cancer detection via mammograms. The CBIS-DDSM (Curated Breast Imaging Subset of Digital Database for Screening Mammography) is one of the key repositories publicly available. It contains 10,239 images and grouped under the labels; Benign, Benign Without Callback and Malignant. Another set of focus is the detection of thoracic conditions via chest x-rays. One publicly available chest x-Ray dataset is CheXpert by the Stanford University School of Medicine. CheXpert contains 224,316 chest radiographs of 65,240 patients. It contains images for 12 different thoracic diseases including Atelectasis, Cardiomegaly, Enlarged Cardiomegaly, Consolidation, Edema, Lung Lesion, Lung Opacity, Pneumonia, Pneumothorax, Fracture, Pleural Effusion and Pleural Other. And it contains 2 other observations "No Finding" and "Support Devices", making 14 observations in total. The radiographs were collected from Stanford Hospital, between October 2002 and July 2017. Another publicly available chest radiograph dataset is MIMIC-CXR dataset by Massachusetts Institute of Technology (MIT). The dataset contains 371,920 chest x-rays associated with 227,943 imaging studies.  Each imaging study contain a frontal view and a lateral view. MIMIC-CXR dataset also contains 14 observations. There is also chest x-ray dataset from the NIH Clinical Center that contains 100,000 x-rays from over 30,000 patients, including many with advanced lung disease. That leads to a total of 696,236 publicly available x-ray images for 12 thoracic conditions.

*Challenges Facing AI In Radiology*

The challenge however lasts in properly testing such systems and ensuring they work in all edge and diverse cases radiologists encounter. A study by Eric Oermann and colleagues found that, deep learning models that detected pneumonia on chest x-rays performed well on further data from sites they were trained on (AUC of 0.93–0.94) but significantly less on external data (AUC 0.75–0.89) (Zech et al., 2018). This demonstrates the challenge of assessing the generality and scalability of Deep Learning systems. Though the study by Liu and colleagues analysed 31,587 studies, only 69 studies provided enough data to construct contingency tables, enabling calculation of test accuracy. And out of that 69 studies, only 25 studies did out-of-sample external validations. And further, only 14 of such studies compared the models' performances to that of radiologists. They also realised the methodology and reporting of studies evaluating deep learning models is variable and often incomplete. This shows the need for standardization of evaluation frameworks and benchmarks for AI radiological systems. This is essential to assessing the quality of Artificial Intelligence solutions, their readiness to be deployed and the degree of autonomy they should be given.

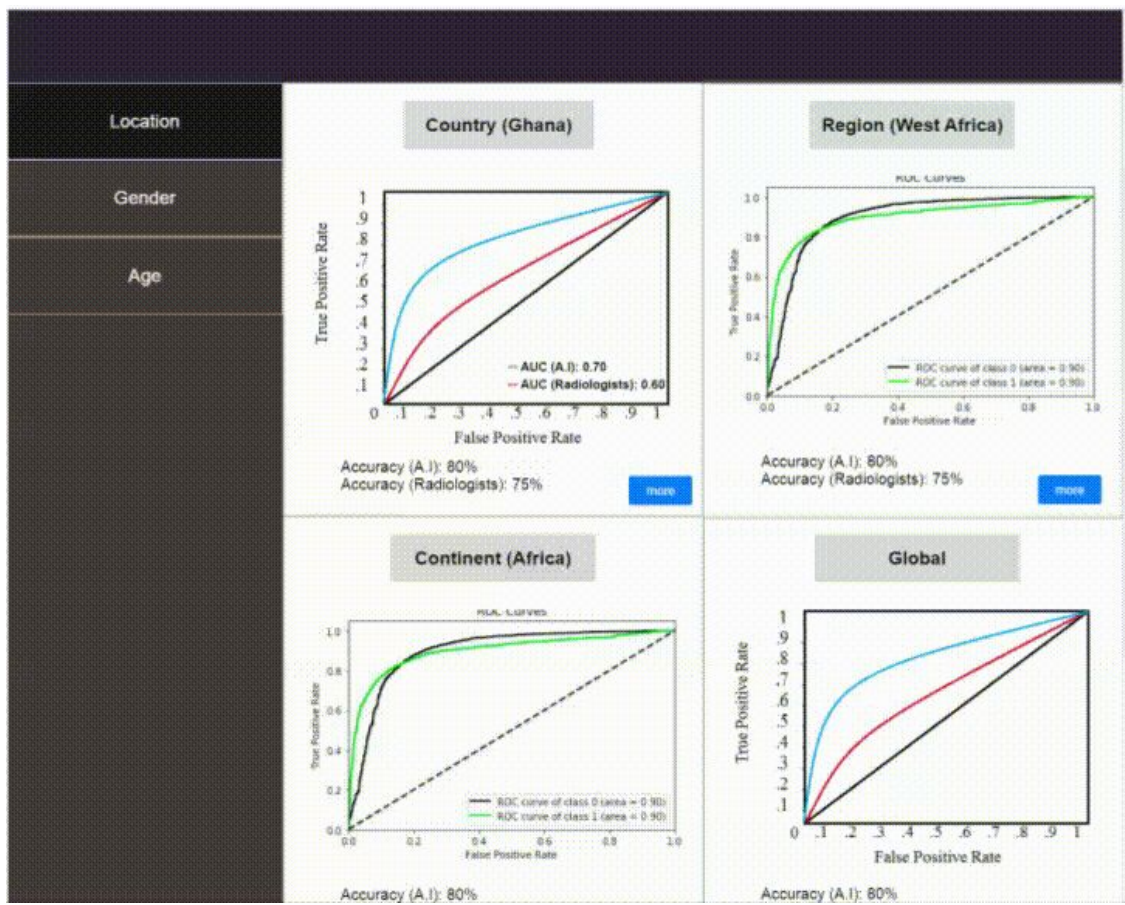

Figure 1: A prototype of the radiograph-agnostic precision evaluation platform.

**Benchmarking Solution**

We are proposing radiograph-agnostic benchmarking platform and framework that would allow for the evaluation of AI radiological systems for various conditions and serve as a standard. This would require registered developers and organisations seeking to evaluate their A.I system to download the test images and a csv file with two columns; 'ID', containing the unique Identification of each test image and 'Class' which would be left blank in order to be populated by the outputs of an A.I system. Developers are then to submit the fully populated csv file, which would then provide the model's outputs to be evaluated with the true labels. Tutorial scripts in popular Machine Learning libraries and frameworks would be provided to developers on how to correctly get your model's outputs to be populated in the csv file.

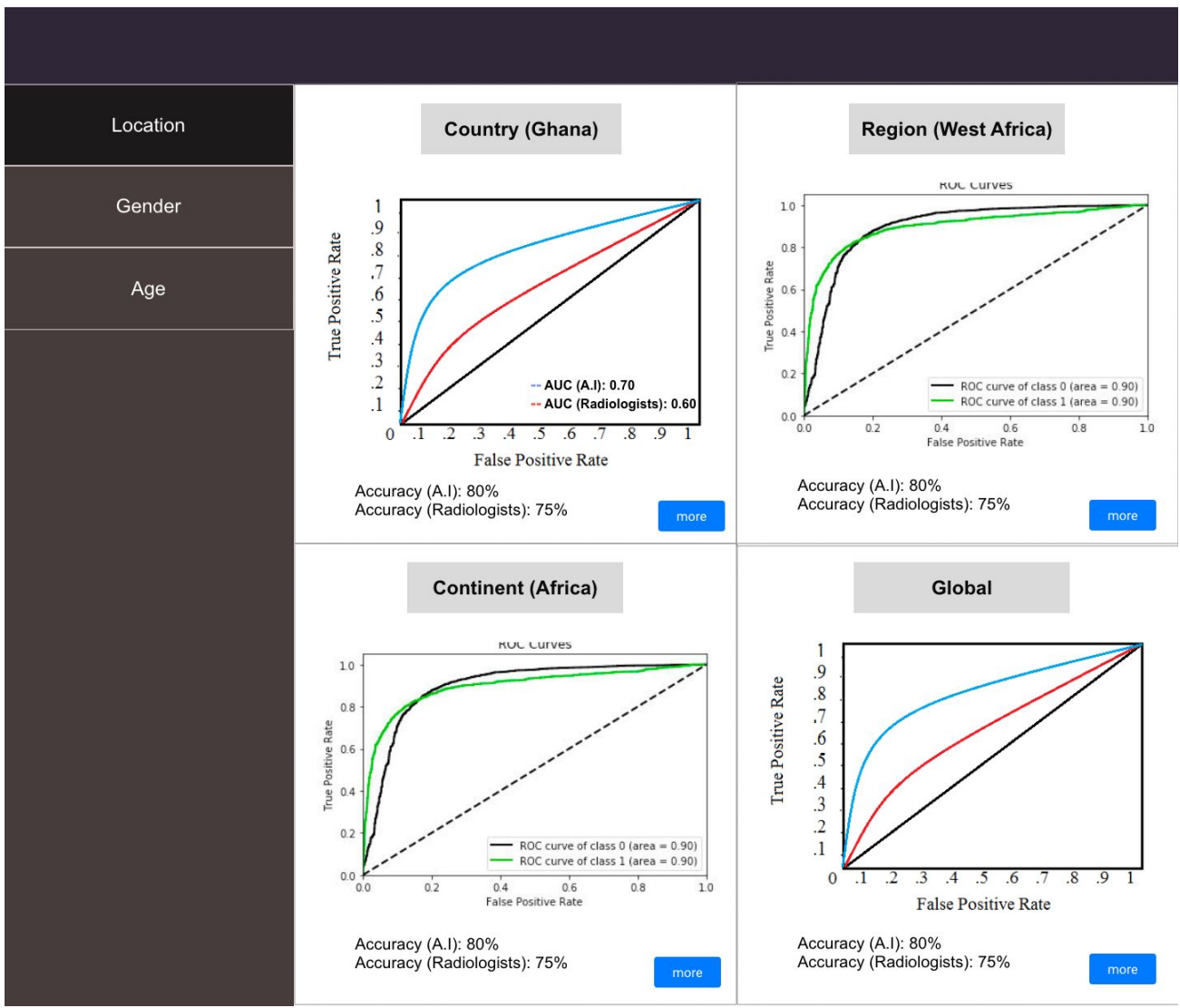

Figure 2: The 'Location' category with its sub-categories and the metrics used.

*Evaluation Metrics*

All our supported condition tests on the platform would be image classification tasks and therefore we would be using evaluation metrics for classification. Some of the conditions and tests would be binary classification tasks while others would be multi-class classification, therefore we would be using metrics that can be used for both types of classification. As shown in Figure 1 and 2, the evaluation metrics to be used would be the Receiver Operating Characteristic (ROC) curve, its Area Under the Curve (AUC) score and the Accuracy Score. The ROC curve and AUC score would help us identify the model's true positive rate (TPR) (Sensitivity) and its false positive rate (FPR) (Specificity). Though originally for binary classification, the ROC curve and AUC score can be generalised to multi-class classification.

The performance of an A.I system would be compared with radiologists using the various metrics. This would help developers see how well their models perform compared to the current popular approach, standalone radiologists. Benchmarking vis-à-vis radiologists would also help in assessing the level of autonomy that should be given each A.I system.

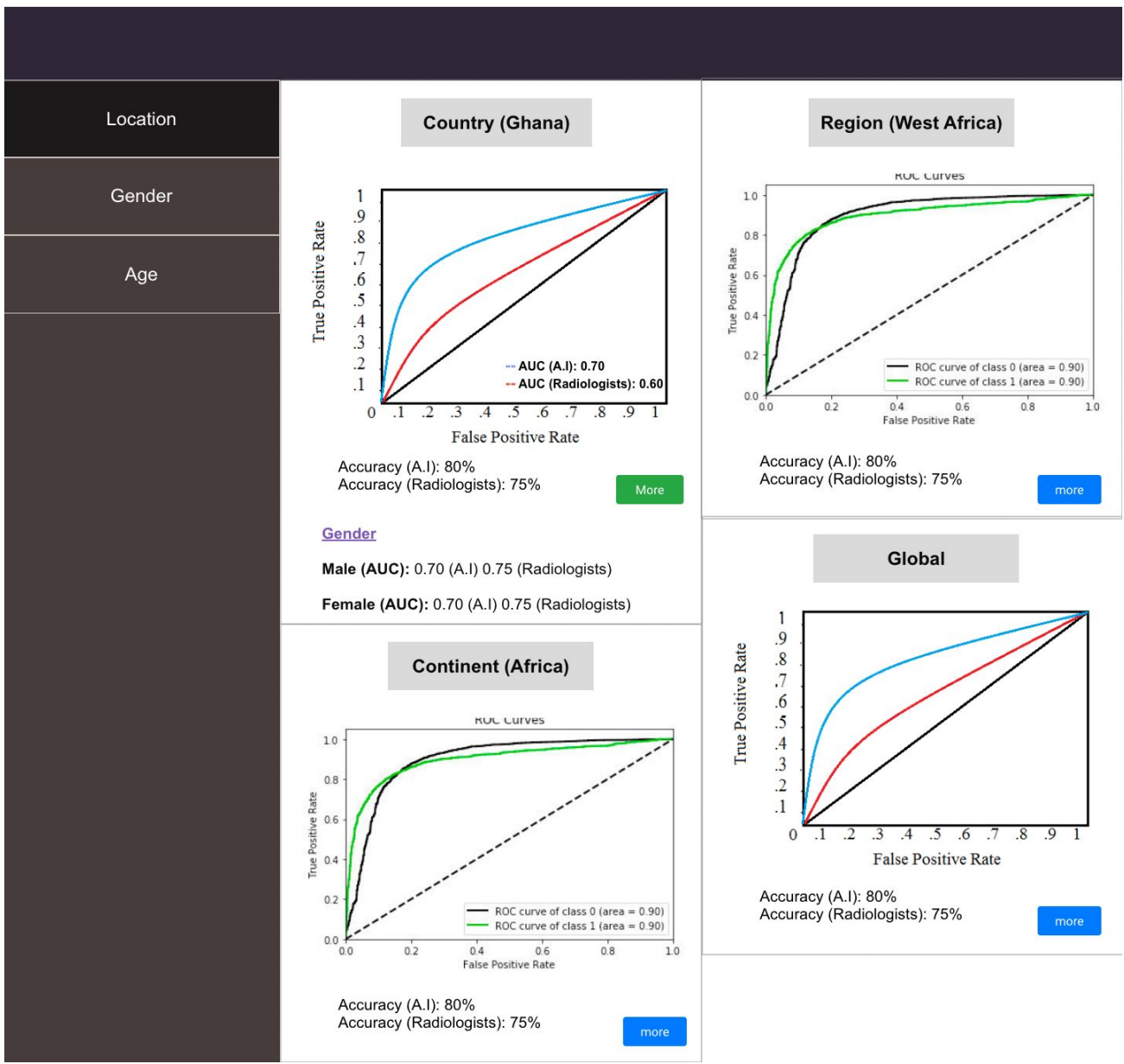

Figure 3: Each sub-category would feature demographics intersection performances too.

*Benchmark Categorizations*

The evaluation results would be divided into Location, Gender and Age, as shown in Figure 1. Under Location, the performance of the AI model would be shown under the sub-categories; Country, Continent, Region and Global. The 'Country' sub-category shows the performance of the A.I system within the very nation it was developed. The 'Continent' sub-category would show how well the model performs on data from the continent it was developed in, this would help the developers know how well they can scale the current version of their A.I system. 'Region' specifically focuses on the performance of the AI system within the sub-continental region it was developed (eg. West Africa, South East Asia, Northern Europe). This would help the developers see how ready their AI system is to be deployed in neighboring countries. And finally, 'Global' shows how well the model performs on data from across the world, showing its ability to truly generalise. Each of the subcategories under location would also feature an AUC score for each Gender and Age group, as shown in Figure 1 and 3. This would allow developers to tell specifically within each geographical area, how well their AI system generalises across gender and age.

Under 'Gender', there would be two main sub-categories, Male and Female, as shown in Figure 1 and 4. This would show how well the AI system performs on radiographs of male and female patients. Each of the two sub-categories would also feature AUC scores for various Age groups. This would show how well the AI system performs on male and female patients of different age groups. Conditions that however only affects one gender would not feature the 'Gender' category.

The 'Age' category would feature various age groups as sub-categories. Age groups that aren't featured within certain datasets and conditions would not be shown for those specific conditions. Similar to the other categories, an AI system's performance on each of the age groups would be shown and it'd also feature 'Male' and 'Female' AUC score under each age group.

This concept of 'Precision Evaluation' is to precisely assess how well an AI system generalises across demographics.

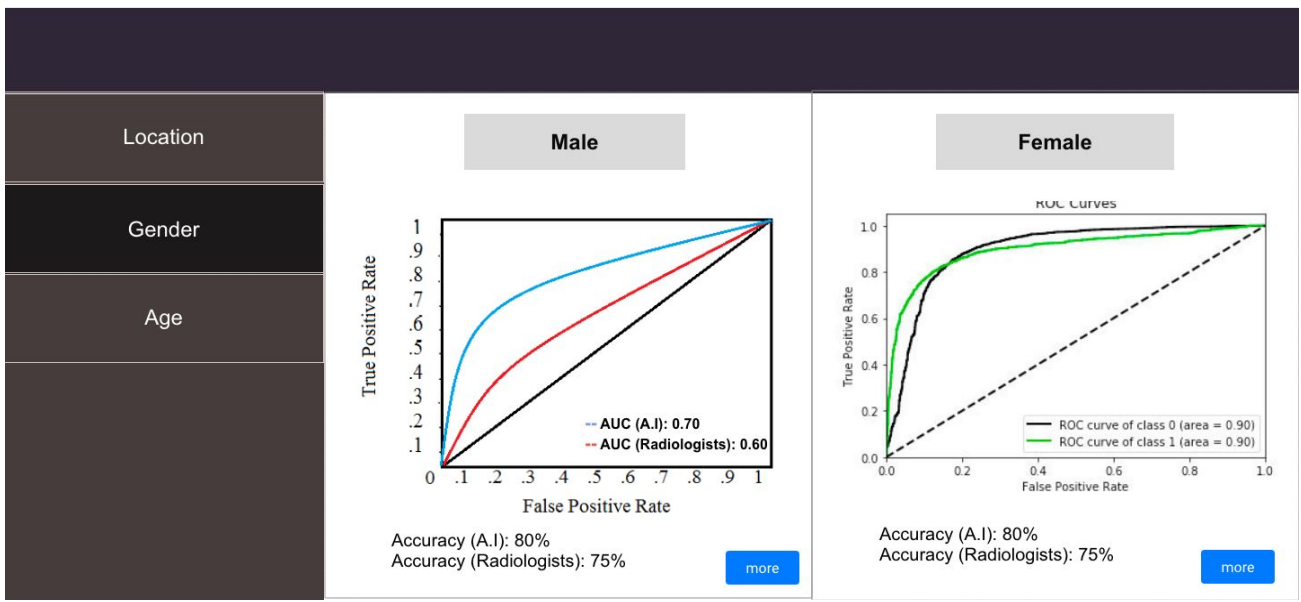

Figure 4: The 'Gender' category.

*Evaluation Data*

The goal is to ensure proportional amount of the diverse demographics and their intersections. With diverse evaluation data, the generality of an AI system can truly be assessed. The platform would be open to facilities to register, and submit images and demographical data. Facilities with approved images would be credited with contributing to the set up of such dataset. This would hopefully serve as incentive to facilities to contribute more data to the platform. Submitted radiographs should be accompanied by a csv file with information about patient's gender, age and imaging facility's location. This would allow for the proposed Precision Evaluation framework.

*The Panel of Expert Radiologists*

To ensure quality, submited images and data would be reviewed by a panel of expert radiologists. This panel of expert radiologists would also ensure edge cases and diversity are represented in each evaluation set. The panel would be open to qualified radiologists to join and participate in. Each evaluation set and condition would have its own panel of experts radiologists. Radiologists who are part of the panel would be credited on the platform for the evaluation sets they contribute to. This would also hopefully serve as an incentive for more radiologists to join 'The Panel of Expert Radiologists'.

*Test Radiologists*

Beyond the panel of expert radiologists, we would ideally have radiologists from different parts of the world who would be asked to classify the test images without access to their true labels. The goal would be to get as many testing radiologists as possible from each continent, region or possibly country. These radiologists would also be ideally given test images from within their region. This would allow us to compare an A.I system's performance on test images within each of the 'Location' sub-category with radiologists also within such geographical region. This would more appropriately help us estimate how well an AI system performs when compared with the level of performance of standalone radiologists within each specific region.

**Evaluation Data Availability**

minoHealth AI Labs is currently working with institutions in Ghana, including Christian Health Association of Ghana (CHAG), National Catholic Health Service (NCHS), Euracare Advanced Diagnostic Center and Paradise Diagnostic Center in order to collect mammograms and chest radiographs. Some of that data can be made available to the benchmarking platform. With the collaboration of various members and organisations affiliated with FG-AI4H, we can collect more radiographs from around the world. Also as explained earlier, the platform would be open to registered facilities to contribute data.

**Feasibility**

Though the proposed radiograph-agnostic framework and platform has several moving parts and complexities, it's possible to modularise it and build with different levels of complexities. It is also possible for the categories and subcategories to adjust based on the number and diversity of samples as well as radiologists available. If the evaluation data for a particular condition isn't large enough to support all four subcategories of 'Location', it can be limited to just 'Region' or 'Continent' and 'Global'. If there weren't enough test radiologists within a specific country where an AI system was developed, the regional, continental or global average performance of radiologists would be used across. The same can apply to the sub-categories of Gender and Age. We would also start implementing the platform with chest x-rays for 12 different thoracic diseases supported in MIMIC-CXR, CheXpert and NIH Chest XRay datasets.

**Privacy and Security**

Anonymised data can be de-anonymised using techniques like linkage attacks. Linkage attacks involves combining data from multiple sources in order to form a whole picture about targets. It is then possible to use the demographics data (Date of Birth, Gender and Location) of an anonymised patient whose medical image is available and cross-reference with public voter lists in order to identify who the patient is. This is because there are very few individuals likely to have the same data of birth and gender, and live in the same location. To prevent linkage attacks, the developers and testing radiologists are only given access to test images without demographics data. To further defend against this attack, we are abstracting 'Date of Birth' to just the Age (in years) of the patient when they were imaged, and we can abstract location to just 'Country'. To add additional security measures as far as the panel of expert radiologists has access to such demographics data, we can explore variations of Differential Privacy.

Also, we are ensuring a secure system by demanding that developers and organisations that require a standardised evaluation of their A.I systems register before they'd be allowed to. The registration process can including an in-person assessment by their local World Health Organisation (W.H.O) or ITU branch office, just to ensure they are a valid institution, startup or developer. A moderate fee can be charged for the registration, which could then serve as funds to support the maintenance of the platform. Equally, health facilities seeking to donate medical images and data must register and be assessed. And even the images and data they submit to the platform would be evaluated before being added to the system. All radiologists, both in the 'panel of expert radiologists' and the 'testing radiologists' would have to register and be verified before allowed to contribute to the platform.

In order to not infringe upon the Intellectual Properties (IP) rights of AI developers and organisations, they would not be required to submit their A.I system itself. They are only supposed to submit the outputs (csv file) of their AI system, which would then be used for the evaluation of their system.

**Impact**

There exists a large amount of publicly available medical image datasets online, and there have been a lot of research and development with such datasets. By developing frameworks that target these conditions first, we would make the standardized benchmarking platform immediately appealing to the A.I healthcare research and development community. This would also help speedup the deployment of AI solutions in Radiology globally. AI healthcare system developers and organisations usually have to go through the challenge of convincing health facilities to share their private data with them, such data unfortunately aren't always of high quality and they usually lack the broad demographic representations needed to truly assess how well an A.I system generalises. A radiograph-agnostic benchmarking platform with data from various facilities across the globe, reviewed by a panel of experts to ensure quality and diversity, would drastically simplify the evaluation stage of such AI systems. The 'Precision Evaluation' framework would help fight against demographically biased A.I systems by ensuring they are tested in great detail across various groups. It'd also help in the safe scaling of AI systems across different locations. The 'Location' sub-categorization of evaluation allows for 'Geo-Precision Evaluation'. Developers can tell how well their systems can perform within their country or first-point of deployment, and should they intend to scale to neighboring countries then eventually have it across the globe, they can tell how well their current version would perform at each point of such growth and scaling.

______________________________